\documentclass[prl,twocolumn,showpacs,preprintnumbers,amsmath,amssymb]{revtex4}
\usepackage{graphicx}   
\usepackage{bm}     
\begin{document}


\title{Exact quantum jump approach to open systems
       in Bosonic and spin baths}

\author{Heinz-Peter Breuer}
\affiliation{Institut f\"ur Physik, Carl von Ossietzky Universit\"at,
         D-26111 Oldenburg, Germany}
\affiliation{Physikalisches Institut, Universit\"at Freiburg,
         D-79104 Freiburg, Germany}
\date{\today}

\begin{abstract}
A general method is developed which enables the exact treatment of
the non-Markovian quantum dynamics of open systems through a Monte
Carlo simulation technique. The method is based on a stochastic
formulation of the von Neumann equation of the composite
system and employs a pair of product states following a Markovian
random jump process. The performance of the method is illustrated
by means of stochastic simulations of the dynamics of open systems
interacting with a Bosonic reservoir at zero temperature and with
a spin bath in the strong coupling regime.
\end{abstract}


\maketitle

The dynamics of open systems plays an important role in a wide
variety of applications of quantum mechanics. The central goal of
the physical theory \cite{TheWork} is to derive tractable
equations of motion for the reduced density matrix $\rho_S(t)$
which is defined by the trace over the environmental variables
coupled to the open system. On the ground of the weak-coupling
assumption the dynamics may be formulated in terms of a quantum
dynamical semigroup which yields a Markovian master equation in
Lindblad form \cite{ALICKI}. However, strong couplings or
interactions with low-temperature reservoirs give rise to large
system-environment correlations which generally result in a
failure of the Markov approximation: The open system dynamics
develops long memory times and exhibits a pronounced non-Markovian
behaviour. Important examples of non-Markovian quantum phenomena
are provided by the atom laser \cite{SAVAGE}, the non-perturbative
decay of atoms coupled to structured reservoirs \cite{GARRAWAY},
environment-induced decoherence at low temperatures, and systems
interacting with a spin bath \cite{STAMP}.

The non-Markovian time-development of open quantum systems
represents a challenge for the analytical and the numerical
treatment, since it requires to cope with a complicated
integro-differential equation \cite{NZ}, or with a highly
non-local influence functional \cite{FEYNMAN}. In the Markovian
regime, Monte Carlo wave function techniques have been shown to
provide efficient numerical tools \cite{SWFM}. In these techniques
one propagates an ensemble of stochastic state vectors $\psi(t)$
in the open system's state space such that the reduced density
matrix is recovered through the ensemble average or expectation
value $\rho_S(t)={\mathrm{E}}(|\psi(t)\rangle\langle\psi(t)|)$. In
view of the numerical efficiency of the method a generalization to
the non-Markovian regime is highly desirable. One such
generalization \cite{DGS} is based on a complicated stochastic
integro-differential equation, involving a time-integration over
the past history of the system dynamics. As shown in
Ref.~\cite{BKP} one can avoid the solution of non-local equations
of motion by propagating a {\textit{pair}} $\psi_1(t)$,
$\psi_2(t)$ of stochastic state vectors and by representing the
reduced system's density matrix through the expectation value
$\rho_S(t)={\mathrm{E}}(|\psi_1(t)\rangle\langle\psi_2(t)|)$. A
similar idea has been used recently to design exact diffusion
processes for systems of identical Bosons \cite{CARUSO} and
Fermions \cite{CHOMAZ}. However, this method demands that one
first derives an appropriate time-local non-Markovian master
equation for the reduced system dynamics. Although this is indeed
possible by use of the time-convolutionless (TCL) projection
operator technique, the derivation becomes extremely complicated
in the strong coupling regime which requires to take into account
higher orders of the TCL expansion of the master equation. Another
possibility \cite{STOCK} is to exploit an explicit expression for
the influence functional of the system. This method is, however,
restricted to Gaussian reservoirs and linear dissipation.

In this Letter an alternative approach to the open system dynamics
is developed which is based on a stochastic formulation of the von
Neumann equation for the density matrix $\rho(t)$ of the
{\textit{total}} system. It is demonstrated that the motion of the
state of any composite quantum system can be described through a
Markovian random jump process which directly translates into a
Monte Carlo simulation technique for the full non-Markovian
reduced system behaviour. By contrast to previous methods which
mainly deal with Bosonic reservoirs and linear dissipation, the
present technique is generally applicable and does not require a
specific form of the interaction between system and environment,
nor any assumption on the physical properties of the environment.

In the interaction picture the Hamiltonian describing the
system-environment coupling can be written as
\begin{equation} \label{H-INT}
 H_I(t) = \sum_{\alpha} A_{\alpha}(t) \otimes B_{\alpha}(t),
\end{equation}
where $A_{\alpha}(t)$ and $B_{\alpha}(t)$ are system and
environment operators, respectively. The dynamics of the total
density matrix is then governed by the von Neumann equation,
\begin{equation} \label{NEUMANN}
 \frac{d}{dt} \rho(t) = -i [H_I(t),\rho(t)].
\end{equation}
Our aim is to express $\rho(t)$ through the mean value
\begin{equation} \label{RHO-REP}
 \rho(t) = {\rm{E}}(|\Phi_1(t)\rangle\langle\Phi_2(t)|).
\end{equation}
$|\Phi_1(t)\rangle$ and $|\Phi_2(t)\rangle$ provide a pair of
state vectors of the combined system which follow an appropriate
stochastic time-evolution. These state vectors are taken to be
direct products of system states $\psi_{\nu}$ and environment
states $\chi_{\nu}$,
\begin{equation} \label{DIRECT-PRODUCT}
 |\Phi_\nu(t)\rangle = \psi_{\nu}(t) \otimes \chi_{\nu}(t),
 \qquad \nu = 1,2.
\end{equation}
Such a representation is indeed possible for any initial state
$\rho(0)$ of the total system, regardless of the physical
structure of open system and environment. In particular, $\rho(0)$
needs not be an uncorrelated state. Equations (\ref{RHO-REP}) and
(\ref{DIRECT-PRODUCT}) amount to representing the density matrix
of the open system through the expression
\begin{equation} \label{RHOS}
 \rho_S(t) = {\mathrm{tr}}_{\mathrm{B}} \rho(t)
 = {\mathrm{E}} \left(
 |\psi_1(t)\rangle\langle\psi_2(t)|
 \langle \chi_2(t) | \chi_1(t) \rangle \right),
\end{equation}
where ${\mathrm{tr}}_{\mathrm{B}}$ denotes the trace over the
variables of the environment. Contrary to the standard method,
expression (\ref{RHOS}) is an average over the product of
$|\psi_1\rangle\langle\psi_2|$ and the overlap $\langle \chi_2 |
\chi_1 \rangle$ of the corresponding environment states.

The task is to construct equations of motion for $\psi_{\nu}$ and
$\chi_{\nu}$. We suppose that the dynamics represents a piecewise
deterministic process (PDP). This is a Markovian stochastic
process whose realizations consist of smooth deterministic pieces
broken by instantaneous jumps occurring at random moments
\cite{DAVIS}. A convenient way of formulating a PDP is to write
stochastic differential equations for the random variables. In our
case an appropriate set of stochastic differential equations is
given by
\begin{eqnarray}
 d\psi_{\nu} &=& \sum_{\alpha}
 \left(\frac{-i||\psi_{\nu}||}{||A_{\alpha}\psi_{\nu}||}A_{\alpha}-I\right)
 \psi_{\nu} dN_{\alpha\nu}, \label{STOCH1} \\
 d\chi_{\nu} &=& \Gamma_{\nu} \chi_{\nu} dt + \sum_{\alpha}
 \left(\frac{||\chi_{\nu}||}{||B_{\alpha}\chi_{\nu}||}B_{\alpha}-I\right)
 \chi_{\nu} dN_{\alpha\nu}, \label{STOCH2}
\end{eqnarray}
where $I$ denotes the unit operator and
\begin{equation} \label{DEFGANU}
 \Gamma_{\nu} \equiv \sum_{\alpha} \Gamma_{\alpha\nu}
 \equiv \sum_{\alpha}
 \frac{||A_{\alpha}\psi_{\nu}||\cdot||B_{\alpha}\chi_{\nu}||}
 {||\psi_{\nu}||\cdot||\chi_{\nu}||}.
\end{equation}
Equations (\ref{STOCH1}) and (\ref{STOCH2}) relate the stochastic
increments $d\psi_{\nu}\equiv\psi_{\nu}(t+dt)-\psi_{\nu}(t)$ and
$d\chi_{\nu}\equiv\chi_{\nu}(t+dt)-\chi_{\nu}(t)$ to a set of
random integers $dN_{\alpha\nu}(t)$ which are known as Poisson
increments and satisfy the relations
\begin{eqnarray}
 dN_{\alpha\nu}(t) dN_{\beta\mu}(t) &=&
 \delta_{\alpha\beta} \delta_{\nu\mu} dN_{\alpha\nu}(t),
 \label{DNDN} \\
 {\mathrm{E}}(dN_{\alpha\nu}(t)) &=& \Gamma_{\alpha\nu}dt.
 \label{EDN}
\end{eqnarray}
Equation (\ref{DNDN}) states that $dN_{\alpha\nu}(t)$ takes on the
possible values $0$ or $1$, while Eq.~(\ref{EDN}) tells us that
$dN_{\alpha\nu}(t)=1$ with probability $\Gamma_{\alpha\nu}dt$.
Moreover, under the condition that $dN_{\alpha\nu}(t)=1$ for a
particular $\alpha$ and $\nu$, all other Poisson increments
vanish. According to Eqs.~(\ref{STOCH1}) and (\ref{STOCH2}) the
state vectors then perform the instantaneous jumps
\[
 \psi_{\nu} \rightarrow
 \frac{-i||\psi_{\nu}||}{||A_{\alpha}\psi_{\nu}||} A_{\alpha}\psi_{\nu},
 \qquad
 \chi_{\nu} \rightarrow
 \frac{||\chi_{\nu}||}{||B_{\alpha}\chi_{\nu}||} B_{\alpha}\chi_{\nu}.
\]
We observe that these jumps conserve the norm of the state vectors
and occur at a rate $\Gamma_{\alpha\nu}$ defined in
Eq.~(\ref{DEFGANU}). Under the condition that all Poisson
increments vanish, that is $dN_{\alpha\nu}(t)=0$ for all $\alpha$,
$\nu$, we have $d\psi_{\nu}=0$ and
$d\chi_{\nu}=\Gamma_{\nu}\chi_{\nu}dt$. Consequently $\psi_{\nu}$
remains unchanged during $dt$, while $\chi_{\nu}$ follows a linear
drift. Summarizing, $\psi_{\nu}(t)$ is a pure, norm-conserving
jump process, whereas $\chi_{\nu}(t)$ is a PDP with
norm-conserving jumps and a linear drift.

The stochastic differential equations (\ref{STOCH1}) and
(\ref{STOCH2}) give rise to an equation of motion for the random
matrix $R(t) = |\Phi_1(t)\rangle\langle\Phi_2(t)|$ whose
expectation value is required to represent the total density
matrix (see Eq.~(\ref{RHO-REP})). To find this equation one
applies the calculus of PDPs, which is analogous to the Ito
calculus of the classical theory of Brownian motion. The
calculations are particularly simple in the present case since
$dN_{\alpha\nu} dN_{\beta\mu}=0$ for $\nu \neq \mu$ (see
Eq.~(\ref{DNDN})). This implies that the increments
$|d\Phi_1\rangle$ and $|d\Phi_2\rangle$ are independent which
yields $dR=|d\Phi_1\rangle\langle\Phi_2|+|\Phi_1\rangle\langle
d\Phi_2|$. Substituting the expressions (\ref{DIRECT-PRODUCT}) and
using the stochastic differential equations (\ref{STOCH1}) and
(\ref{STOCH2}) as well as the relations (\ref{DNDN}) and
(\ref{EDN}) one finds
\begin{equation} \label{DR}
 dR(t) = -i [H_I(t),R(t)]dt + dS(t),
\end{equation}
where $dS=dT_1R+RdT_2^{\dagger}$ is a stochastic increment and
\[
 dT_{\nu} = \sum_{\alpha} \left(
 \Gamma_{\alpha\nu} dt - dN_{\alpha\nu} \right)
 \left(I+i\Gamma^{-1}_{\alpha\nu} A_{\alpha}B_{\alpha} \right).
\]
Since the quantity $\Gamma_{\alpha\nu} dt - dN_{\alpha\nu}$ is
zero on average by virtue of Eq.~(\ref{EDN}), we conclude that
${\mathrm{E}}(dT_{\nu})=0$ and, hence, ${\mathrm{E}}(dS)=0$. The
role of the drift contributions $\Gamma_{\alpha\nu}dt$ is to
compensate the mean contributions ${\mathrm{E}}(dN_{\alpha\nu})$
of the jumps towards the stochastic increment $dS$. The average
over the stochastic equation (\ref{DR}) is therefore identical to
the von Neumann equation (\ref{NEUMANN}). This shows that the PDP
defined by Eqs.~(\ref{STOCH1}) and (\ref{STOCH2}) indeed
reproduces the exact von Neumann dynamics of the combined system
through the expectation value (\ref{RHO-REP}).

The random process thus constructed leads to a Monte Carlo
simulation technique of the open system dynamics. The numerical
algorithms that can be used to generate a sample of realizations
of the PDP are similar to the ones employed in the standard
unraveling of quantum Markovian master equations. The decisive
difference is, however, that the present method works with a pair
of stochastic states and uses a representation in terms of states
of the total system. Additionally, these features allow the
determination of all kinds of multi-time quantum correlation
functions directly from the Monte Carlo technique.

An important limitation of the Monte Carlo method is set by the
size $\sigma(t)$ of the statistical fluctuations. A detailed
analysis of the stochastic differential equations reveals that
$\sigma(t)$ is bounded for any finite time $t$, but may eventually
grow exponentially at a rate of at most $2\Gamma_0$, where
$\Gamma_0$ is an upper bound for the transitions rates
$\Gamma_{\nu}$. The stochastic simulation technique is thus useful
for short and intermediate times, the relevant time scale being
given by $1/2\Gamma_0$. It is impossible, in general, to use the
scheme for large times, trying to beat an exponential increase of
the fluctuations by enlarging the number of realizations. However,
the statistical errors may be considerably reduced by employing
the fact that the $|\Phi_{\nu}\rangle$ evolve independently, or by
using a class of stochastic states with a more complicated
structure.

To illustrate the method we first study the model of a two-state
system with excited state $|e\rangle$, ground state $|g\rangle$,
and transition frequency $\omega_0$. This system is coupled to a
Bosonic reservoir consisting of a continuum of field modes $k$
with creation and annihilation operators $b^{\dagger}_k$, $b_k$,
and corresponding frequency $\omega_k$. The interaction
Hamiltonian is taken to be
\begin{equation} \label{H-I-BOSONIC}
 H_I(t) = \sigma_+ B(t) + \sigma_- B^{\dagger}(t),
\end{equation}
with the system operators $\sigma_+=|e\rangle\langle g|$ and
$\sigma_-=|g\rangle\langle e|$ and the reservoir operator
$B(t)=\sum_k g_k b_k e^{i(\omega_0-\omega_k)t}$. The $g_k$ are
mode-dependent coupling constants. We investigate the dynamics
evolving from the initial state
$|\Phi_{\nu}(0)\rangle=\psi_{\nu}(0)\otimes\chi_{\nu}(0)=|e\rangle\otimes
|0\rangle$, where $|0\rangle$ denotes the vacuum state of the
reservoir. The central quantity that determines the influence of
the reservoir modes on the reduced system dynamics is provided by
the correlation function $\langle 0|B(t')B^{\dagger}(t)|0\rangle$
which may be expressed in terms of the reservoir spectral density
$J(\omega)$.

\begin{figure}[htb]
\includegraphics[width=\linewidth]{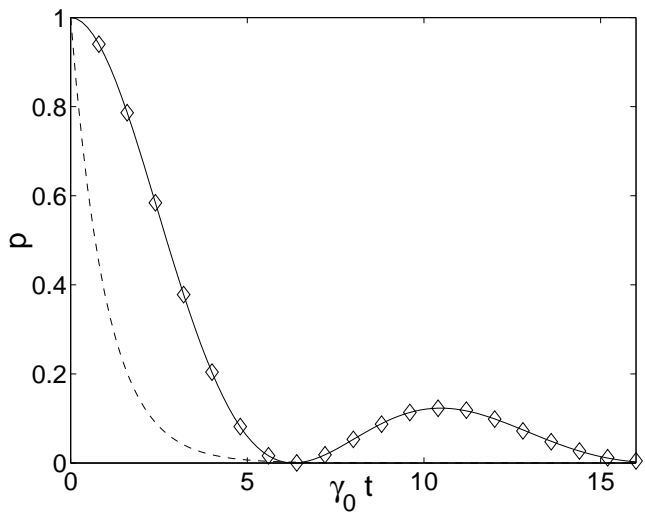}
\caption{\label{figure1}
 The excited state probability $p(t)$ of a two-state system
 coupled to a Bosonic reservoir:
 Analytical solution (continuous line) and Monte Carlo simulation
 of Eqs.~(\ref{STOCH1}), (\ref{STOCH2}) (diamonds) for
 $\gamma_0/\lambda = 5$ with $2\cdot 10^7$ realizations.
 The broken line shows the Born-Markov approximation.}
\end{figure}

Figure \ref{figure1} shows results of a Monte Carlo simulation of
the stochastic differential equations (\ref{STOCH1}) and
(\ref{STOCH2}) for the damped Jaynes-Cummings model defined by the
spectral density
$J(\omega)=\frac{1}{2\pi}\gamma_0\lambda^2/[(\omega_{0}-\omega)^2+\lambda^2]$.
The figure displays the population $p(t)=\langle
e|\rho_S(t)|e\rangle$ of the excited state, estimated with the
help of expression (\ref{RHOS}). For the parameter values chosen
the reservoir correlation time $\lambda^{-1}$ is five times larger
than the reduced system's Markovian relaxation time
$\gamma_0^{-1}$. This shows that the case under study belongs to
the strong coupling regime and that substantial deviations from
the Born-Markov dynamics occur, as is indicated in
Fig.~\ref{figure1}.

For small and intermediate couplings the reduced system dynamics
is known to satisfy a non-Markovian master equation with a
time-dependent generator, which can be obtained with the help of
the TCL expansion in the coupling constant $\gamma_0/\lambda$.
However, the resulting perturbation expansion of the master
equation diverges in the strong coupling regime
$\gamma_0/\lambda>0.5$ for times which are larger than the first
positive zero $t_0$ of $p(t)$. Beyond this singularity the
generator of the master equation does not exist. The TCL master
equation is therefore not capable of describing the revival of the
excited state population for $t>t_0$. By contrast, the stochastic
representation developed here reproduces the exact evolution of
the reduced system and, hence, correctly describes the dynamics
even beyond the singularity.

The stochastic simulation method can also be applied to the
dynamics of open systems interacting with a spin bath, a quantum
environment whose properties differ substantially from those of a
Bosonic reservoir. We examine a central spin model consisting of a
spin with Pauli spin operator $\vec{\sigma}$ which is coupled to a
bath of $N$ spins described by spin operators
$\vec{\sigma}^{(j)}$, $j=1,2,\ldots,N$. The interaction
Hamiltonian reads
\begin{equation} \label{HINT-SPIN}
 H_I(t) = \sigma_3 B_3 + \sigma_+ B_-(t) + \sigma_- B_+(t),
\end{equation}
where $B_{\pm}(t)=\sum_j2A^{(j)}\sigma_{\pm}^{(j)}e^{\mp
i\omega_0t}$, $B_3=\sum_jA^{(j)}\sigma_3^{(j)}$. This model may be
used to describe the interaction of a single electron spin,
confined to a quantum dot, with an external magnetic field and a
bath of nuclear spins through hyperfine interactions \cite{LOSS}.
The transition frequency of the central spin is $\omega_0$ and we
set $A^{(j)}=A/\sqrt{N}$, such that $A$ provides the root mean
square of the couplings of the central spin to the various bath
spins.

To give an example of a mixed initial state we choose
$\rho(0)=|+\rangle\langle-|\otimes 2^{-N}I$, where $|\pm\rangle$
are eigenstates of the 3-component $\sigma_3$ of the central spin
and $I$ is the unit matrix in the state space of the bath.
Initially, the bath is thus in a completely unpolarized state.
This state can efficiently be realized through a mixture of basis
states which are simultaneous eigenstates of the square of the
total spin angular momentum $\vec{J}$ of the bath and of its
3-component $J_3$, with an appropriate distribution of the
corresponding quantum numbers $j$ and $m$. Employing an argument
given in \cite{BOSE} we conclude that the probability of finding
the quantum numbers $(j,m)$ in the initial mixture can be written
as
$P(j,m)=2^{-N}\left[\binom{N}{N/2+j}-\binom{N}{N/2+j+1}\right]$.
This method of generating the initial bath state bears the
advantage that it enables one to employ the conservation of the
3-component $\frac{1}{2}\sigma_3+J_3$ of the total spin angular
momentum and to carry out a canonical transformation which removes
the term $\sigma_3 B_3$ in the interaction Hamiltonian
(\ref{HINT-SPIN}). This is an example of the implementation of
known symmetries and conservation laws within the simulation
algorithm.

\begin{figure}[htb]
\includegraphics[width=\linewidth]{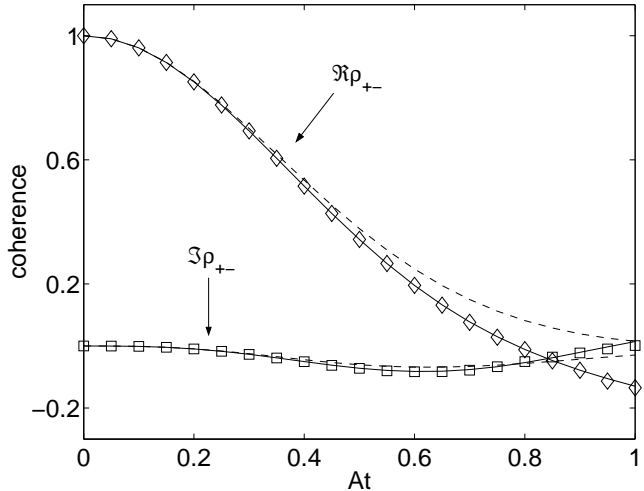}
\caption{\label{figure2}
 Real part $\Re\rho_{+-}$ and imaginary part $\Im\rho_{+-}$ of the
 coherence of a central spin coupled to a
 bath of $10^3$ spins through the Hamiltonian (\ref{HINT-SPIN})
 with $A/\omega_0=0.5$. Symbols:
 Simulation of Eqs.~(\ref{STOCH1}), (\ref{STOCH2}) with
 $2\cdot 10^7$ realizations. Continuous lines: Solution of the
 von Neumann equation (\ref{NEUMANN}). Broken lines:
 TCL master equation in second order.}
\end{figure}

Figure \ref{figure2} shows the coherence
$\rho_{+-}(t)=\langle+|\rho_S(t)|-\rangle$ of the central spin
obtained from a Monte Carlo simulation employing again expression
(\ref{RHOS}) for the reduced density matrix. To asses the
performance of the stochastic method we compare the simulation
results with the solution of the von Neumann equation for the
total system. The simulation of the PDP obviously reproduces the
von Neumann dynamics with high accuracy, the statistical errors of
the simulation being smaller than the size of the symbols. For the
parameter values chosen, the dynamics significantly deviates from
the predictions of the second order TCL master equation.

The stochastic formulation naturally lends itself to several
interesting generalizations. One possibility is to formulate the
dynamics in imaginary time by means of the substitution
$t\rightarrow -i\beta$, to examine the equilibrium properties of
the system ($\beta$ denotes the inverse temperature). With slight
modifications of the differential equations (\ref{STOCH1}) and
(\ref{STOCH2}) one finds a stochastic dynamics in imaginary time
which unravels the canonical density matrix of the system via the
expectation value
$\rho(\beta)={\mathrm{E}}(|\Phi_1(\beta)\rangle\langle\Phi_2(\beta)|)$.

It was assumed in Eq.~(\ref{DIRECT-PRODUCT}) that the
$|\Phi_{\nu}\rangle$ are tensor products of state vectors of
system and environment. Since the interaction creates
system-environment correlations it may be advantageous to employ a
certain class of {\em{entangled}} stochastic states with the aim
of a more efficient representation of $\rho(t)$ as mean value over
the underlying random process. A further possibility is to use a
stochastic propagation of {\textit{mixed}} states. This is
achieved, for example, by the introduction of a stochastic matrix
$R(t)=|\psi_1(t)\rangle\langle\psi_2(t)|\otimes
R_{\mathrm{B}}(t)$, where the $\psi_{\nu}$ are states of the open
system and $R_{\mathrm{B}}$ is a random operator in the state
space of the environment. It is then again possible to construct
stochastic differential equations leading to the exact von Neumann
dynamics with the help of the mean value
$\rho(t)={\mathrm{E}}(R(t))$. A systematic exploration of these
ideas could be of great relevance for the development of efficient
numerical algorithm of the quantum dynamics of open systems.


\begin{thebibliography}{99}
\bibitem{TheWork} H. P. Breuer and F. Petruccione,
                  \textit{The Theory of Open Quantum Systems}
                  (Oxford University Press, Oxford, 2002).
\bibitem{ALICKI} R. Alicki and K. Lendi, \textit{Quantum Dynamical
                 Semigroups and Applications}, Lecture Notes in Physics 286
                 (Springer-Verlag, Berlin, 1987).
\bibitem{SAVAGE} G. M. Moy, J. J. Hope, and C. M. Savage,
                 Phys. Rev. A \textbf{59}, 667 (1999);
                 J. J. Hope, G. M. Moy, M. J. Collett, and C. M. Savage,
                 Phys. Rev. A \textbf{61}, 023603 (2000).
\bibitem{GARRAWAY} B. M. Garraway, Phys. Rev. A \textbf{55}, 2290 (1997);
                   Phys. Rev. A \textbf{55}, 4636 (1997).
\bibitem{STAMP} N. V. Prokof'ev and P. C. E. Stamp,
                Rep. Prog. Phys. \textbf{63}, 669 (2000).
\bibitem{NZ} S. Nakajima, Progr. Theor. Phys. \textbf{20}, 948
             (1958); R. Zwanzig, J. Chem. Phys. \textbf{33}, 1338 (1960).
\bibitem{FEYNMAN} R. P. Feynman and F. L. Vernon,
                  Ann. Phys. (N.Y.) \textbf{24}, 118 (1963);
                  A. O. Caldeira and A. J. Leggett,
                  Physica \textbf{121A}, 587 (1983).
\bibitem{SWFM} J. Dalibard, Y. Castin, and K. M{\o}lmer,
               Phys. Rev. Lett. \textbf{68}, 580 (1992);
               R. Dum, P. Zoller, and H. Ritsch,
               Phys. Rev. A \textbf{45}, 4879 (1992);
               N. Gisin and I. C. Percival, J. Math. Phys. A: Math. Gen.
               \textbf{25}, 5677 (1992);
               H. Carmichael, \textit{An Open Systems Approach
               to Quantum Optics}, Lecture Notes in Physics m18
               (Springer-Verlag, Berlin, 1993);
               M. B. Plenio and P. L. Knight,
               Rev. Mod. Phys. \textbf{70}, 101 (1998);
               A. Imamoglu, Phys. Rev. A {\textbf{50}}, 3650
               (1994).
\bibitem{DGS} W. T. Strunz, L. Di\`osi, and N. Gisin,
              Phys. Rev. Lett. \textbf{82}, 1801 (1999).
\bibitem{BKP} H. P. Breuer, B. Kappler, and F. Petruccione,
              Phys. Rev. A \textbf{59}, 1633 (1999).
\bibitem{CARUSO} I. Carusotto, Y. Castin, and J. Dalibard,
                 Phys. Rev. A \textbf{63}, 023606 (2001);
                 I. Carusotto and Y. Castin,
                 J. Phys. B: At. Mol. Opt. Phys. \textbf{34}, 4589 (2001).
\bibitem{CHOMAZ} O. Juillet and Ph. Chomaz,
                 Phys. Rev. Lett. \textbf{88}, 142503 (2002).
\bibitem{STOCK} J. T. Stockburger and H. Grabert,
                Phys. Rev. Lett. \textbf{88}, 170407 (2002).
\bibitem{DAVIS} M. H. A. Davis, \textit{Markov Models and Optimization}
                (Chapman \& Hall, London, 1993).
\bibitem{LOSS} A. V. Khaetskii, D. Loss, and L. Glazman,
               Phys. Rev. Lett. \textbf{88}, 186802 (2002).
\bibitem{BOSE} A. Hutton and S. Bose, quant-ph/0208114.
\end{thebibliography}
\end{document}